# Sentient Networks

George Chapline
Lawrence Livermore National Laboratory
chapline1@llnl.gov

Abstract

The engineering problems of constructing autonomous networks of sensors and data processors that can provide alerts for dangerous situations provide a new context for debating the question whether man-made systems can emulate the cognitive capabilities of the mammalian brain. In this paper we consider the question whether a distributed network of sensors and data processors can form "perceptions" based on sensory data. Because sensory data can have exponentially many explanations, the use of a central data processor to analyze the outputs from a large ensemble of sensors will in general introduce unacceptable latencies for responding to dangerous situations. A better idea is to use a distributed "Helmholtz machine" architecture in which the sensors are connected to a network of simple processors, and the collective state of the network as a whole provides an explanation for the sensory data. In general communication within such a network will require time division multiplexing, which opens the door to the possibility that with certain refinements to the Helmholtz machine architecture it may be possible to build sensor networks that exhibit a form of artificial consciousness.

## 1. Introduction

During the next few decades autonomous networks of sensors and data processors are going to come into widespread use for industrial process monitoring and detection of threats to populations and infrastructures. If the individual sensors have sufficient sensitivity and the environmental data are unambiguous, then these sensor networks can be operated as robots; which means the sensory data can be used to directly activate responses. Obviously automatic warning systems that can quickly respond to malfunctions in an industrial process or hostile acts which threaten populations or essential infrastructures would have many practical applications. Unfortunately, in practice it may be very difficult on the basis of outputs from sensors to decide in a timely way that something untoward is happening. For example, it might be that state of the art sensors do not have sufficient sensitivity or discrimination to provide unambiguos signals regarding malfunctions or hostile acts. Even in networks where the individual detectors are very sensitive it may be necessary to correlate the outputs from sensors in different locations and also possibly from different kinds of sensors in order to determine whether a dangerous situation has arisen. Another frequent problem that occurs with the interpretation of real world sensor data is that signal to noise ratios for the signatures of interest are not sufficiently large to provide unambiguous signals.

These problems are similar to those faced by many species of animals in the natural world. Actually almost all animals can respond to dangers. However, in all invertebrates (with the possible exception of cephalopods) this response is always a pre-conditioned automatic response. For example in order to navigate around obstacles the brains of flying insects must almost instantaneously recognize how these obstacles are moving relative to the insect, and then immediately send appropriate signals to the muscles controling its flight. It is the goal of most current robotics research projects to duplicate this kind of automatic response in various kinds of man-made electromechanical devices. In vertebrates, on the other hand, there is a new anatomical feature - the cerebral cortex - which provides an additional channel for responding to sensory inputs. As a consequence vertebrates can respond to dangers either by making use of a rapid midbrain response system similar to that in invertebrates or in a somewhat slower fashion by making use of their cerebral cortex. In the cerebral cortex of vertebrates sensory data is processed in a much more sophisticated way than in the brains of invertebrates. Not only is the extraction of features from the sensory data more elaborate, but studies in both humans and animals of the functions of the cerebral cortex suggest that the processing of sensory data in the cerebral cortex is primarily concerned with the formation of explanations for the sensory data; i.e. "perceptions" [1].

One of the obvious evolutionary advantages of being able to take into account various possible explanations for sensory data is that this greatly enhances an animal's capability to cope with stealthy predators. A vivid example of the necessity of having such a capability is provided by the dilemma faced by a zebra on the African savanah. A lion sitting in plain view is very likely not to be an immediate threat, whereas recognizing that one is being stalked by a lion will often require making that determination on the basis of manifestly ambiguous data - for example, a slight movement of the grass or the state of agitation of birds or other animals. It is clear that faced with ambiguous information an animal may or may not want to respond. Indeed it is tempting to speculate that vertebrates employ some kind of statistical inference engine in order to determine whether in ambiguius situations there is a danger that requires flight or some other prompt respone. In fact there is strong circumstancial evidence [2] that humans in effect make use of a statistical inference system in order to make decisions. Although initially it might seem implausible that humans - not to mention animals with no obvious capabilities for understanding abstract concepts - could utilize probabilistic reasoning, it is perhaps logically inescapable [3] that the vertebrate cerebral cortex must be capable of inferring and somehow encoding estimates of the posterior probabilities for the occurrence of various situations or events that are important for survival. Historically it was suggested more than a century ago by Helmholtz that the main function of the human perceptual system is to infer probable causes for sensory inputs.

In the following section 2 we briefly review the well known statistical basis for the assertion that it is the availability of posterior probabilities that makes it possible for a network of sensors and processors to provide explanations for sensory data. Unfortunately it is unknown exactly how the vertebrate cerebral cortex evaluates and records posterior probabilites. Nevertheless some parallel computation architectures have appeared in the literature which, at least in principle, should make it possible to use posterior probabilities to provide explanations for sensory data. Although parallel computation architectures for the interpretation of sensor outputs have not yet come into widespread use in engineering systems, there are reasons for expecting that this type of architecture will find widespread use in the future when real time responses are desired. For example, even in the case of robotic feedback control systems where the use of parallel computation architectures may not be mandatory the significant speed advantages of parallel analog circuits will make the use of VLSI parallel analog circuits [4] increasingly attractive for the real time interpretation of sensor data [5]. In the case of distributed networks of sensors rapid interpretation of the sensory data will almost certainly require new computational paradigms. In particular, when the sensory data are ambiguous, the use of parallel computation is almost manditory because in general there can be exponentially many explanations for a particular set of sensor outputs.

The rationale for parallel computation in those cases where one is faced with an exponential proliferation of possible explanations for the sensor outputs is that the required computational task of finding the most likely explanation is intractible for a single sequential data processor. This will be especially true when finding an explanation for sensor data involves using Markov chain Monte Carlo methods to "invert" a stochastic model for the world. On the other hand, as noted by Hopfield and Tank in a celebrated paper [6], parallel computation using a "neural network" architecture provides a remarkable capability for finding good solutions to computationally intractible problems. Consequently although various kinds of computational schemes, e.g. rule based expert systems, might be employed in special cases to rapidly explain sensory data, artificial neural networks do provide a natural universal computational framework to use in an autonomous system seeking an explanation for sensor data precisely because of their ability to provide good solutions to computationally difficult problems. Furthermore although there are various parallel computation architectures that might be adopted to the problem of explaining sensor data we would like to focus in the following on the fact that physical systems of interacting 2-state units, viz. "spins", in thermal equilibrium with a heat bath can duplicate the behavior of artificial neural networks like those introduced by Hopfield, and in fact have certain advantages over completely deterministic neural networks (this is the basic idea behind simulated annealing [7]).

The main point we would like to make in this paper is that in a distributed network of autonomous sensors and data processors alternative interpretations of sensor inputs can be represented by the collective state of physically separated binary units, each of which has activation level 0 or 1. In other words an autonomous network of sensors and data processors can be conceived of as the hardware embodiment of an abstract statistical system of interacting spins functioning as a neural network. In section 4 we describe one particular way, which we call the *Helmholtz machine* after the corresponding abstract neural network architecture, in which a hardware network of sensors and data processors might be made to emulate a spin system functioning as a statistical inference engine . In section 5 we briefly discuss how a hardware Helmholtz machine might be built using existing data processing and communication technologies. In particular we compare and contrast the intelligent agents required for a Helmholtz machine with the TCP/IP interfaces used in Ethernet networks. In section 6 we address the question whether a distributed network of sensors and simple data processors could behave as though it were "conscious".

## 2. Statistical theory of pattern recognition

It has been understood for some time that pattern recognition systems are in essence machines that utilize either preconceived probability distributions or empirically determined posterior probabilities to classify patterns. In the ideal case where the a priori probability distribution $p(\alpha)$ for the occurence of various classes $\alpha$ of feature vectors and probability densities $p(x|\alpha)$ for the distribution of data sets x within each class are known, then the best possible classification procedure would be to simply choose the class $\alpha$ for which the posterior probability

$$P(\alpha \mid x) = \frac{p(\alpha)p(x \mid \alpha)}{\sum_{\beta} p(\beta)p(x \mid \beta)} \tag{1}$$

is largest. Unfortunately in the real world one is typically faced with the situation that neither the class probabilities $p(\alpha)$ nor class densities $p(x|\alpha)$ are precisely known, so that one must rely on empirical information to estimate the conditional probabilities $P(\alpha \mid x)$ needed to classify data sets. In practice this means that one must adopt a parametric model for the class probabilities and densities, and then use empirical data to fix the parameters $\theta$ of the probability model. Once values for the model parameters have been fixed, then sensory data can be classified by simply substituting values for the model probabilities $p(\alpha; \theta)$ and $p(x|\alpha; \theta)$ into equation (1).

Unfortunately determining values for the model parameters from empirical data is itself a computationally intractible problem. This means that in practice one is usually limited to using models of relatively modest complexity, and consequently one is always faced with the issue of choosing the best possible values for the model patameters. One popular way of measuring how good a paticular set of model parameters is at reproducing the observed data, known as the maximum likelihood (ML) estimator, can be motivated by noting that the formula for the posterior probability given in equation (1) can be formally interpreted as the canonical Boltzmann distribution for the population of energy levels of a physical system in equilibrium with a heat bath. In particular if one defines the "energy" of a classification $\alpha$ to be

$$E_\alpha = -\log p(\alpha) p(x \mid \alpha), \qquad (2)$$

then the posterior probability introduced in (1) can be formally expressed in the form

$$P(\alpha \mid x) = \frac{e^{-E_\alpha}}{\sum_\alpha e^{-E_\alpha}}, \qquad (3)$$

where the energies are those defined in equation (2). If we suppose that the energy levels $E_\alpha$ defined in equation (2) define a physical system whose energy levels are populated according to a canonical distribution of the form (3), then the thermodynamic free energy of this system will be given by

$$F(x) = \sum_\alpha \{E_\alpha P(\alpha) - (-P(\alpha) \log P(\alpha))\}, \qquad (4)$$

where we have used $P(\alpha)$ as shorthand notation for the canonical distribution (3). If instead of the true probability distributions p($\alpha$) and p(x|$\alpha$) one uses model probabilities p($\alpha$; $\theta$) and p(x|$\alpha$; $\theta$) to calculate a probability distribution $P_\theta(\alpha)$ for different classifications of a data set x, then equation (3) will no longer necessarily be satisfied and the free energy calculated from eqation (4) will in general differ from the true free energy. In particular we would have that

$$F(x) = F(x, \theta) - \sum_\alpha P_\theta(\alpha) \log[P_\theta(\alpha) / P(\alpha)] \qquad (5)$$

where $F(x, \theta)$ is the free energy calculated using the distribution $P_\theta(\alpha)$. The quantity $\sum_\alpha P_\theta(\alpha) \log[P_\theta(\alpha) / P(\alpha)]$ in the second term in equation (5) is always positive and measures of the difference in bits between the model distribution $P_\theta(\alpha)$ and the true distribution $P(\alpha)$. This distance measure, known as the *Kullback- Leibler divergence,* is the basis for the ML estimator that is widely used by statisticians to measure how well a given set of model probabilities reproduces the empirical data [8]. Minimization of the Kullback-Leibler distribution with respect to the model parameters $\theta$ is one way of choosing the best possible values for these parameters. It should also be noted that the estimated free energy $F(x, \theta)$ is always greater than the equilibrium free energy $F(x)$, so that the best possible probability distribution to use for classifications is that which minmizes the estimated free energy. This analogy between pattern recognition and statistical physics opens the door to using insights from theoretical physics to solve pattern recognition problems.

An ingenious physics model in which the posterior probabilities $P(\alpha | x; \theta)$ are naturally represented in the canonical Boltzmann distribution form (3) was introduced in 1985 by Ackley, Hinton, and Sejnowski [9]. In this model, known as the *Boltzmann machine*, data sets and their "explanations" are represented by configurations of binary units with activation levels $a_i = 0$ or 1. The energy function for the assembly of binary units is assumed to have a form similar to that used by physicists to describe a system of interacting spins in a magnet:

$$E(\mathbf{a}) = -\frac{1}{2}\sum_{i \neq j} w_{ij} a_i a_j + \sum_i \theta_i a_i, \qquad (6)$$

where $\mathbf{a} = \{a_i\}$ denotes the set of activation levels, the $\theta_i$ are biases, and the weights $w_{ij}$ describes the interaction strength between binary units i and j In the work of Ackley et. al. these interactions are assumed to be symmetric; i.e. $w_{ij} = w_{ji}$. If the system of binary units is assumed to be in contact with a heat bath at some fxed temperature the probabilty distribution for the activation levels will approach a stationary equilibrium whose form is just the Boltzmann distribution (3) corresponding to the energy function (6). In the case of the Boltzmann machine the probability distribution $P_\theta(\alpha)$ will be the probability distribution for the activation levels in a certain subset, referred to as the hidden units, of all binary units. The remaining binary units, referred to as the visible units, represent the environmental data x. The model parameters $\theta$ for the Boltzmann machine are the connection strengths $w_{ij}$ and biases $\theta_i$ for the binary units. These parameters are determined by minimizing the Kullbach-Leibler divergence between the probability distribution $P_\theta(\alpha)$ with the visible unit activation levels fixed and the probability distribution for classifications with the activation levels of the visible units allowed to vary freely. Used as a pattern recognition device the Boltzmann machine has the virtue that non-trivial correlations between different instances of environmental data are automatically represented, and used in the classification of data sets. This means that the classifications provided by the Boltzmann machine take into account more information than just the relationship between a class and its feature vectors. Unfortunately Boltzmann machines have not found many practical applications because determination of the connection strengths and biases for realistic data sets is very slow using even the fastest supercomputers. Moreover even when the connection strengths and biases are known the classification of data sets is slow because of the necessity for repetitive sampling of a joint probability distribution for the activation levels of the hidden units.

## 3. Belief networks

As noted in the introduction vertebrates are often faced with the problem of finding an explanation for sensory data whose interpretation is not immediately obvious. This is a rather more subtle problem than the more familiar problem of classifying static patterns because finding an explanation for sensory data will in general involve taking into account structural relationships that are inherent in the entire ensemble of environmental input data. One of the main goals of artificial intelligence research is the development of algorithms which can recognize and manipulate the structural relationships inherent in data sets. Almost effortlessly the human brain is able to capture the essence of these structural relationships by the use of hierarchical knowledge trees or "ontologies". At the present time there is a great deal of interest within the artificial intelligence community in developing algorithmic tools for contructing and utilizing ontologies for applications such as medical diagnosis and battlefield management [10]. It remains unclear though whether these purely algorithmic approaches to artificial intelligence will lead to systems that can deal with the ambiguities and complexities of real world sensor data.

In existing computerized decision systems which use probabilistic reasoning the knowledge base is represented by a description of the dependencies between the variables of the system [11]. In a *belief network* these dependencies are represented by a probability distribution for the states of the nodes in a layered network which is a product of conditional probabilities for each unit given the values of the units which proceed it in some ordering. Many types of belief networks are possible, but typically belief networks have a tree-like structure and the conditional probabilities have the Markov property; i.e. the random variables at nodes not connected by a branch are conditionally independent given those variables which are so connected. In addition belief networks often incorporate unobserved latent variables known as hidden variables. Software implementations of Markov belief networks with hidden variables have been successfully used for some quite difficult real world pattern recognition problems such as speech recognition [12], and represent a promising general approach to the interpretation of complex data sets. Indeed the success of hidden variable Markov models for speech recognition should certainly be kept in mind when designing networks of sensors and data processors to interpret environmental data, and is also perhaps a significant clue as to how the cerebral cortex of vertebrates forms perceptions .

Models for the cerebral cortex which might function as belief networks were first introduced by Little, Shaw, and Vasudevan [13], and indeed these models were among the first artificial neural networks. The nodes of these networks can be represented by binary units with activation level $a_i$ =0 or 1, and it has been pointed out [13] [i] that if the Markov transition probability for the activation level of a single node in these models have the sigmoidal form

$$p(a_i(n+1) \mid \mathbf{a}(n)) = \sigma[\beta(1-2a_i(n+1))\sum_{j} w_{ij}a_j], \tag{7}$$

where $\sigma(x) = 1/[1+\exp(-x)]$ then these models resemble Boltzmann machines. The vector $\mathbf{a}(n) = \{a_i(n)\}$ in equation (7) denotes the set of activation levels at layer n of the network. In the original work of Little, Shaw, and Vasudevan the way activation levels vary from layer to layer was thought of as describing the time evolution of the whole network; whereas in their use as belief networks the number n will generally be taken to represent actual physical layers of the network. A little algebra shows that the Markov transition probability for the ensemble of activation levels in a layer can be written in the form

$$P[\mathbf{a}(n+1) \mid \mathbf{a}(n)] = \frac{e^{\beta a(n+1) \bullet X(n)}}{\sum_{a(n+1)} e^{\beta a(n+1) \bullet X(n)}}, \tag{8}$$

where $X_i(n) = \sum_{j} w_{ij}a_j(n) - \theta_i a_i$ and we have used vector multiplication notation in the exponents. If the interactions between units are symmetric, then the transition matrix (8) will eventually become independent of n, and the probability distribution for activation levels will have the canonical Boltzmann form; if activation levels in the first layer are fixed we recover equation (3). If the connection strengths between units are asymmetric, i.e. $w_{ij} \neq w_{ji}$, then the activation levels will vary from layer to layer and depend on the activities in the first layer. The activation levels in the initial layer represent the environmental data, while the activation levels in later layers can be thought of as representing the explanation of the the environmental data [15]. It should be noted in this connection that 2-dimensional arrays of binary units within each layer provide a convenien way of representing and classifying feature vectors using redundant population codes [16]. Thus Boltzmann machines with asymmetric weights provide an attractive way to realize fault tolerent belief networks.

One of the main practical problems with using conventional belief networks to explain sensor data is that given a set of conditional probabilities and associated decision tree which constitutes a model for the world, finding explanations for input data will in typically involve using Markov chain Monte Carlo methods to invert the world model. Because of the need for repetitive sampling, this cannot in general be done in real time. Another practical problem with belief networks is that determining the network parameters from the environmental data is very difficult if there are too many parameters. Fortunately a descendent of the the Boltzmann machine form of belief networks, the *Helmholtz machine*, has recently emerged that offers the promise of alleviating these problems.

## 4. Helmholtz machine networks

The Helmholtz machine architecture introduced by Hinton et.al. [17] is a stochastic neural network that resembles the Boltzmann machine in that the nodes of the network consist of units whose activation levels are quantized to be either 0 or 1. Another similarity between the Helmholtz and Boltzmann machines is that some of the binary units represent environmental input data, while the remaining "hidden units" represent possible explanations for the input data. All information concerning conditional probabilities is contained in the values of the connection strengths $w_{ij}$ between nodes; indeed it is possible that these weights play much the same role in the Helmholtz machine as the synaptic connection in the cerebral cortex. The activation of the ith hidden unit in the network is chosen stochastically in accordance with the probability $p_j(x) = p(a_i(n+1) | \mathbf{a}(n))$ that with a particular set $x$ of sensory inputs the ith hidden unit in the nth layer has activation 1. The probabilities $p_j(x)$ are calculated from the weights for the connection between the ith unit and the activities of the units in the previous layer using the sigmoidal formula (7). Thus the activities of units are determined in a quasi-deterministic way similar to that in a feed-forward neural network. If one assumes that the activities of the binary units within a given layer are independent, then the probability of a particular explanation α ={$\mathbf{a}$(n), n>1} will be given by the product :

$$Q(\alpha) = \prod_{n>1} \prod_j [p(a_j)]^{a_j} [1 - p(a_j)]^{1-a_j} ; \qquad (9)$$

so that the binary units that are turned on contribute with weight $p_j(x)$ while the units that are turned off contribute with weight 1- $p_j(x)$.

All of the information concerning the structure of the external world that is needed to explain a given ensemble of sensory data is encoded into the connection weights $w_{ij}$ used in eq. 6 , and the main obstacle in utilizing a Helmholtz machine is determining the values of these parameters. Unfortunately when the set of input data is complex it is a computationally intractible problem to precisely determine these parameters. However, one can obtain approximate values for the connection weights of a Helmholtz machine based on the observation (cf. eq's 3 and 5) that the negative free energy -$F$ (x, θ, Q ) calculated using the non-equilibrium distribution Q(α) provides an lower bound for the logarithm of the probability of generating a particular set x of sensory inputs. In the Helmholtz machine scheme of Hinton et. al. the parameters θ correspond to the connection weights and biases of a separate feed forward network which is used offline to generate the "true" posterior probability distribution $P_\theta(\alpha)$. The connection strengths $w_{ij}$ which are used in the recognition network to explain input data are determined simultaneously with the parameters θ by using gradient descent methods to minimize the free energy function F(x, θ, Q ). Because the conditioning of almost any kind of network that used to interpret observed data will be computationally tedious, it is perhaps to be expected that artificial neural networks will increasingly be used for the purpose of determining the parameters of belief networks; and the use of a separate generative network for this purpose by Hinton et. al. may be a first step in this direction.

Once the connection strengths of the Helmholtz machine have been fixed, then the probabilities for likely explanations of given sensory data can be rapidly determined using Eq's 7 and 8. Thus the Helmholtz machine architecture avoids the problem with the Boltzmann machine of having to repetitively sample a probability distribution in order to recognize input data. Furthermore with the use of the independence ansatz, eq. 8, this architecture allows one to sidestep the combinatorial problem that arises when there are exponentially many explanations. In summary, in addition to making the learning of the structure inherent in a set of input patterns tractable, the separate generative and recognition models used in a Helmholtz machine allow one to arrive at definitive interpretations of instances of environmental data in real time.

Of course from the perspective of actually building sensor networks the crucial feature of the Helmholtz machine concept is that it can apparently be realized in a straightforward way in hardware as a distributed system of sensors and simple data processors. In addition it would be natural in such a distributed system to use population codes to represent both feature vectors and possible explanations of the environmental data, thus decentralizing the cognitive process of the machine. Once the connection strengths for each individual hidden unit are determined they can be stored at the location of the hidden unit and used in a simple data processor to stochastically determine the activation level of that particular hidden unit in response to sensory inputs. It should be emphasized that a very nice feature of this way of realizing the Helmholtz machine architecture is that it is very robust against failures of individual components; i.e. a significant number of individual sensors and/or hidden units would have to fail before the system as a whole completely loses its ability to interpret sensory data.

The actual way in which the sensors and hidden units of a Helmholtz machine are deployed for the detection of threats or undesirable situations will depend on the particular application. For example, in the case of a biological warfare warning system, one might be interested in detecting expanding clouds of small particles. For this application the particle size detectors need to be coupled to the hidden units in such a way as to detect the motion and expansion of a cloud of spore sized particles. Traditional methods for confirming the presence of biological pathogens such as immunoassays would be too slow to use in an emergency response system. However, other techniques, e.g. spectral analysis of backscattered UV radiation, might provide sufficient hints which together with the cloud tracking units could be used to infer whether microbiological warfare agents have been released. This example is in fact a good illustration of the kind of situation where a Helmholtz machine could to be used to fuse multi-modality sensor data and compute a probability that a dangerous siuation had arisen even though the data provided by each sensor modality is highly ambiguous. It should be noted that nothing about the physics of how clouds of microbiological agents spread or the physics of remote sensing needs to be understood a priori by the Helmholtz machine. A Helmholtz machine is capable of self-learning, and all that is required to recognize that a biological warfare attack is underway is that the Helmholtz machine be trained to recognize the regularities and structure of the sensory inputs expected for such an attack.

**5. Intelligent agents for sentient networks**

In a Helmholtz machine network both the processed data and computational processes are distributed throughout the network. If the sensor and data processor nodes are physically separated then some means must be provided for these nodes to communicate with each other. This communication support must be capable of relaying the activation levels of the binary units in one level of the network to the units in the next level of the network within a relevancy time interval. Since each node of the network maintains a list of the other nodes it must communicate with, it may be convenient to make use of Ethernet networking technology to implement the Helmholtz machine architecture. Indeed one advantage of configuring the Helmholtz machine as an Ethernet is that one could take advantage of the IEEE standards (1451.1 and 1451.2) for connecting sensors to microprocessors and the sensor plus microprocessor to a network [18]. A significant caveat here though is that in practice it may only be possible to configure the Helmholtz machine as an Ethernet if the nodes can be connected by wires or cables.

In cases where the nodes of the network are geographically separated it may be necessary to use wireless communications; and in these cases it may be more desirable to use some other interface between the sensor plus microprocessor and the network than the TCP/IP agent used in Ethernets. In general though one will want to use time division multiplexing for the communications in order to minimize the total energy expended in network communications. In particular if one uses an internal clock for the network as a whole in which each node is assigned a talk and listen time slot, then all communication devices can be powered down except during those few time slots in which they participate.

The total bandwidth required for the network communications will be determined by the response time desired and the number of nodes that must communicate with each other during a clock cycle. Generally, each hidden processor unit needs to communicate with every binary unit in the previous layer of the network. However, only a small number of bits needs to be transmitted between any two nodes at any given time, and this will reduce local bandwidth requirements. Furtherrmore in many practical cases the connection strengths between distant units may be negligible so that communication need be carried out only between nearby neighbors. Thus it appears that for the network as a whole an Internet-like model for the communications where local networks feed data to hubs which can communicate with each other using much higher bandwidth transmission channels might be appropriate. In any case because of the practical requirement to conserve energy resources asynchronous sampling techniques like those used by pager receivers would probably not be acceptable for the nodes of a distributed Helmholtz machine network. In the next section we suggest that this practical requirement may have some profound consequences for our understanding of the cerebral cortex.

## 6. Artificial consciousness?

If a network of sensors and simple processors configured as a Helmholtz machine might be considered to be sentient , an obvious question is under what conditions would such networks become "conscious"? The neurophysiological meaning of consciousness is still a matter of considerable controversy. Nevertheless following the suggestion by Crick and Koch [19 ] that consciousness is intimately related to the presence in the cerebral cortex of 40 Hz oscillations, there is increasing evidence [20] that coordination of the activities within the cerebral cortex by an internal clock is somehow responsible for the unique awareness of the external world associated with consciousness. In addition an intriguing mathematical interpretation for a correlation between the use of an internal clock in a neural network and conscious awareness has recently emerged [21]; namely construction of a continuous 3-dimensional manifold from a foliation of 2-dimensional representations of feature vectors will not be topologically possible without using an internal clock if the 2-dimensional representations lie on a topologically non-trivial surface.

As noted previously layering in a belief network is essential for accomodating hierarchical explanations of the input data. Furthermore in the case of a Helmholtz machine it is natural to represent these hierarchical features or "interpretations" using 2-dimensional population codes. An interesting question is under what circumstances can a foliation of 2-dimensional population code representations be fused together to form a 3-dimensional manifold? There is compelling evidence that the human brain makes use of continuous 3-dimensional representations of the external world [22], and indeed construction of such representations may be a very nice way of fusing different data modalities in artificial threat detection systems. The interesting topological fact is that this will only be possible if there is an internal clock and each node is assigned a time slot. In fact 2-dimensional data representations will necessarily lie on a topologically non-trivial surface if they are part of a Hopfield-like network. Further this circumstance will occur naturally if the feature vectors are self-organized [21], and in addition Hopfield-like networks could serve the purpose of error correction for feature vectors [23]. An additional benefit [21] of fusing topologically non-trivial representations of feature vectors is that the only stable excited states for the network are those that correspond to excitations on the boundary; i.e. variations in input data. This of course is very reminiscent of the phenomenology of the conscious cerebral cortex. Thus perhaps our most exciting observation is that with some slight modifications the Helmholtz machine architecture might provide a framework for creating artificially consciousness in a network of sensors and data processors.

**Acknowledgements**
I am very grateful to John Woodworth and Alan Spero for encouraging me to think about the collective properties of sensor networks. I would also like to thank Jim Barbieri, Bill Buchanan and Dave Fuess for helpful discusssions. This work was supported under DOE contract W-7405-ENG-48.

**References**

1. A. S. Bregman, "Perceptual Interpretation and the Neurobiology of Perceptions" in *The Mind-Brain Continuum* ed. R. Llinas and P. S. Churchland (MIT Press 1996).

2. D. Kahneman and A. Tversky, Cognitive Psychology $\underline{3}$, 430 (1972).

3. H. Reichenbach, "Predictive Knowledge" in *The Rise of Scientific Philosophy* (University of California Press 1951).

4. G. Chapline, C. Y. Fu, and B. Law, "Neural Networks on Chips" in *Energy and Technology Review Oct.-Dec. 1992* , ed. J. Sefcik (Lawrence Livermore National Laboratory 1992).

5. N. Franceschini, "Engineering Applications of Small Brains" in Future Electron Devices $\underline{7}$, 38 (1996).

6. J. J. Hopfield and D. W. Tank, Biol. Cybern. $\underline{52}$, 141 (1986).

7. S. Kirkpatrick, C. D. Gelatt, and M. P. Veccchi, Science $\underline{220}$, 671 (1983).

8. G. E. Hinton and T. J. Sejnowski, "Learning and Unlearning in Boltzmann Machines" in *Parallel Distributed Processing* , ed. D. E. Rumelhart et. al. (MIT Press, 1986).

9. S. Kullback, *Information Theory and Statistics* (Wiley, 1959).

10. For an introduction to current research in this area see the Web site for Stanford University's Knowledge Systems Laboratory.

11. J. Pearl, *Probabilistic Inference in Intellegent Systems* (Morgan Kaufmann 1988).

12. L. R. Rabiner and B. H. Juang, *Fundamentals of Speech Recognition* (Prentice Hall 1993).

13. W. A. Little, Math. Biosci. 19, 101 (1974); G. L. Shaw and R. Vasudevan, Math. Biosci. 21, 207 (1974).

14. B. Apolloni and D. de Falco, Neural Comp. 3, 402 (1991).

15. R. M. Neal, Neural Comp. 4, 832 (1992).

16. R. S. Zemel and G. E. Hinton, Neural Comp. 7, 549 (1995).

17. P. Dayan, G. Hinton, and R. Neal, Neural Comp. 7, 889 (1995).

18. J. Warrior, "Smart Sensor Networks of the Future" in *Sensors Mar. 1997*, 40.

19. F. Crick and C. Koch, Seminars in the Neurosciences 2, 263 (1990).

20. R. Llinas and D. Pare, "The Brain as a Closed System Modulated by the Senses" in *The Mind-Brain Continuum* ed. R. Llinas and P. S. Churchland (MIT Press 1996).

21. G. Chapline, Network: Comp. Neural Syst. 8, 185 (1997).

22. R. N. Shepard, Psychonomic Bulletin & Review 1, 2 (1994).

23. G. H. Hinton and T. Shallice, Psychol. Rev. 98, 74 (1991).

---

[i] "Cooperative Monitoring Program, Business Plan", US Department of Energy, Office of Nonproliferation and National Security, Office of Research and Development, 1996.